%% file: main.tex
\documentclass[acmsmall,screen,authorversion]{acmart}
\makeatletter
\renewcommand{\@copyrightowner}{N. Namashivayam}
\makeatother

\input{utils/packages}      
\input{utils/def}           

\begin{document}

\title[GPU-centric Communication Schemes for HPC and ML Applications]{Exhibiting
GPU-Centric Communications}
\subtitle{A Survey on Communication Schemes for Distributed HPC and
ML Applications}
\author{Naveen Namashivayam}
\orcid{0000-0003-3695-5696}
\affiliation{%
    \institution{University of Minnesota}
    \country{USA}
}

\begin{abstract}
\input{src/content/abstract}
\end{abstract}



\maketitle

\renewcommand{\shortauthors}{N.Ravi}

\input{src/content/introduction}
\input{src/content/background}
\input{src/content/schemes}
\input{src/content/implement}
\input{src/content/challenges}
\input{src/content/conclusion}

\bibliography{main}

\end{document}

%% file: utils/packages.tex
\usepackage{verbatim}
\usepackage{graphicx}
\usepackage{url}
\usepackage{epstopdf}
\usepackage{float}
\usepackage{graphicx}
\usepackage{wrapfig}
\usepackage{comment}
\usepackage{csquotes}
\usepackage{url}
\usepackage{xcolor,colortbl}
\usepackage{color}
\usepackage{fancyvrb}
\usepackage{epsfig}\restylefloat{table}
\usepackage{setspace}
\usepackage{multirow}
\usepackage{threeparttable}
\usepackage{mathtools}
\usepackage[ruled, linesnumbered, vlined]{algorithm2e}
\usepackage{algpseudocode}\algrenewcommand\textproc{}
\usepackage{listings}
\usepackage{listings}
\usepackage {natbib} 
\usepackage[misc]{ifsym}
\usepackage[nolist]{acronym}
\usepackage{tikz}
\usepackage{flushend}
\usepackage{framed}
\usepackage{subcaption}

%% file: utils/def.tex
\SetAlFnt{\small}
\SetAlCapFnt{\small}
\SetAlCapNameFnt{\small}
\SetAlCapHSkip{0pt}
\IncMargin{-\parindent}

\algdef{SE}[DOWHILE]{Do}{doWhile}{\algorithmicdo}[1]{\algorithmicwhile\ #1}%

\usetikzlibrary{calc}
\makeatletter
\newlength{\mylength}
\xdef\CircleFactor{1.1}
\newsavebox{\mybox}
\newcommand*\circled[2][draw=blue]{\savebox\mybox{\vbox{\vphantom{WL1/}#1}}\setlength\mylength{\dimexpr\CircleFactor\dimexpr\ht\mybox+\dp\mybox\relax\relax}\tikzset{mystyle/.style={circle,#1,minimum height={\mylength}}}
\tikz[baseline=(char.base)]
\node[mystyle] (char) {#2};}
\makeatletter

%% file: src/content/abstract.tex
Compute nodes on modern heterogeneous supercomputing systems comprise CPUs,
GPUs, and high-speed network interconnects (NICs). Parallelization is identified
as a technique for effectively utilizing these systems to execute scalable
simulation and deep learning workloads. The resulting inter-process
communication from the distributed execution of these parallel
workloads is one of the key factors contributing to its performance bottleneck.
Most programming models and runtime systems enabling the communication
requirements on these systems support GPU-aware communication schemes that move
the GPU-attached communication buffers in the application directly from the GPU
to the NIC without staging through the host memory. A CPU thread is required to
orchestrate the communication operations even with support for such
GPU-awareness. This survey discusses various available \textit{GPU-centric
communication} schemes that move the control path of the communication
operations from the CPU to the GPU. This work presents the need for the new
communication schemes, various GPU and NIC capabilities required to implement
the schemes, and the potential use-cases addressed. Based on these discussions,
challenges involved in supporting the exhibited GPU-centric communication
schemes are discussed.

%% file: src/content/introduction.tex
\section{Introduction}\label{src:introduction}
To accommodate efficient inter-process communication in modern heterogeneous 
supercomputing systems~\cite{frontier1,frontier2,perlmutter,lumi,alps,aurora,elcapitan}
comprised of CPUs and GPUs, the distributed simulation and deep learning 
applications executed on these large-scale systems that span across different
application domains such as molecular dynamics~\cite{LAMMPS,QMCpack,GROMACS},
genome analysis~\cite{BerkeleyGW},
quantum field theory~\cite{MILC,LatticeQCD1,LatticeQCD2},
weather modeling~\cite{CosmoFlow},
solid and fluid mechanics~\cite{Nekbone,M-PSDNS},
material modeling~\cite{QuantumEspresso}, astrophysics~\cite{TOAST3},
and artificial neural network training~\cite{DeepCam,LBANN,GPT3}
use \textit{GPU-aware communication} libraries.
GPU-aware libraries~\cite{craympi,mvapich,anlmpich,openmpi,nccl,rccl} support
performing inter-process communication operations~\cite{specmpi,pgasspec}
involving GPU-attached memory buffers without having the application to stage 
them through CPU-attached memory.
Remote direct memory access (RDMA)~\cite{RDMA} and GPU vendor-specific
peer-to-peer~\cite{gpup2p} data transfer mechanisms are used to implement
GPU-aware inter-node and intra-node inter-process data movement operations on
the scale-out~\cite{slingshot,infiniband,efa} and
scale-up~\cite{nvlink,infinity-fabric} interconnects available in the
system, respectively.

Even with the state-of-the-art GPU-awareness in the communication stack, CPU threads
are still required to orchestrate data-moving communication and inter-process
synchronization operations. CPU threads are required to synchronize with the
GPU and NIC to manage this orchestration. While the programming models and
runtimes support multiple effective options (polling, interrupt, or callbacks)
in implementing the required synchronization, it is unnecessary for the CPU to
get involved in the communication path. The reliance of an CPU thread to
orchestrate the communication in these existing state-of-the-art GPU-aware
communication schemes can impact (1) compute/communication overlap, (2)
communication latency, and (3) GPU autonomy. The current communication model
necessitates a rigid workflow and impacts the application's performance
~\cite{takashi-field-phase-simulator,quentin-distributed-transformer,
aashaka-taccl,amedeo-in-network-aggregation,wei-deeplight,
nadeen-in-network-aggregation}.

Modern GPUs and NICs support various capabilities, such as support for
persistent GPU kernels~\cite{persistent-kernels}, communication operations with
deferred execution semantics~\cite{triggered-ops}, NIC-enabled memory
ordering~\cite{gpudirect-async}, and programmable processors on the
NIC~\cite{dpu1,dpu2,dpu3,dpu4}, making it possible to eliminate the CPU from orchestrating the
communication operations. This work surveys the various available
communication schemes to support GPU orchestrating the communication operations.
The new communication schemes observed in this work enable efficient
communication/compute overlap, improved communication latency and GPU autonomy.

\subsection{Contributions}\label{src:contributions}
As described in this work, \textit{GPU-centric communication} represents a
communication operation that allows the GPU to directly manage the data movement
operation involving a GPU-attached memory buffer. Three different GPU-centric
communication schemes are explored in this work - (1) stream triggered
communication, (2) kernel triggered communication, and (3) kernel initiated
communication. Apart from describing the various available GPU-centric
communication options, this work briefly discusses the hardware and software
features required to implement them and the potential challenges involved. To
our knowledge, this is the first survey to comprehensively describe the various
available GPU-centric communication schemes.

\subsection{Related Surveys}\label{src:related}
Other surveys in GPU-based communication focus on parallelism requirements from
the applications~\cite{tal-ben-ddnn,daniel-ddnn,yoshua-dnn-representations}, scalable system architectures
~\cite{benjamin-thesis,massimo-commodity}, and potential programming model
semantics~\cite{pbridges-mpi,taylor-perf-tradeoff} for enabling GPU-centric
communication. Two works~\cite{ismayil-thesis,didem-survey} in particular
attempts to explore the different properties of GPU-centric communication
operations but focuses on a subset of the available options.
Existing surveys focus on a subset of the available GPU-centric
communication operations or implementation strategies. This work comprehensively
analyzes the existing state-of-the-art communication scheme and a complete list
of available GPU-centric options by decoupling the communication schemes from
the implementation and programming model semantics.

\subsection{Scope}\label{src:org}
This work provides a comprehensive analysis of the various GPU-centric
communication schemes. It is organized as follows:
\begin{itemize}
  \item Section~\ref{src:bground} provides the background on the communication
  operations described in this work.
  \item Section~\ref{src:comms-baseline} describes the currently employed
  baseline and state-of-the-art communication schemes used for inter-process
  communication.
  \item Section~\ref{src:schemes} exhibits on the various available GPU-centric
  communication schemes.
  \item Section~\ref{src:impact} demonstrates the various factors impacting the
  communication schemes exhibited in this work and the critical software and
  hardware capabilities required for implementing them.
  \item Section~\ref{src:apps} describes the various communication patterns
  addressed by the GPU-centric communication schemes, and
  \item Section~\ref{src:challenges} extrapolates potential challenges involved
  in implementing the discussed schemes and provide concluding remarks in
  Section~\ref{src:conclude}.
\end{itemize}

%% file: src/content/background.tex
\section{Background}\label{src:bground}
\begin{figure}[!ht]
  \centering
  \includegraphics[width=\linewidth]{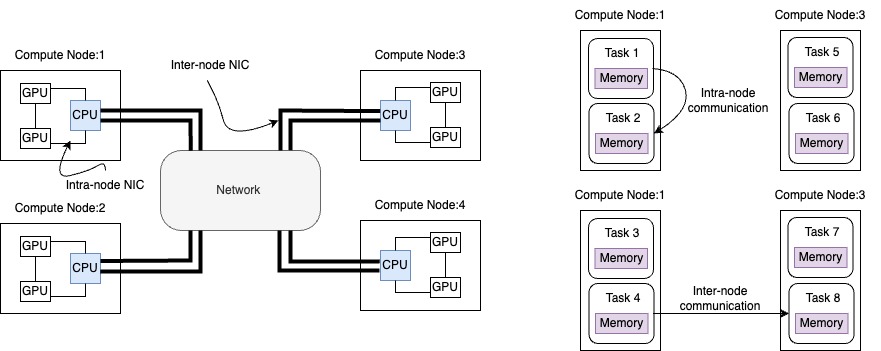}
  \caption{Representing a traditional HPC heterogeneous system architecture with
  four compute nodes connected across a network. The heterogeneous compute nodes
  represent a host CPU attached to two GPU devices. Eight tasks are created for
  the distributed application, each placed on the same compute node.}
  \label{fig:hetero-systems}
\end{figure}

This section describes the communication operations classified in this work. An
arbitrary HPC system used in this work consists of the following components, as
shown in Fig.~\ref{fig:hetero-systems}: (1) heterogeneous compute nodes with
a host processor (CPU), accelerator units (like GPUs), and memory systems,
(2) intra-node or scale-up network connecting the various compute units inside a
compute node, and (3) inter-node or scale-out network connecting the various
compute nodes in the system.

The host and accelerator units in the compute nodes have a shared or independent
memory subsystem based on the compute node architecture. This section briefly
describes the compute node architecture to provide more details on the resulting
communication operations executed from the nodes. Multiple contemporary compute
node architectures are discussed in Section~\ref{src:systems}.

Similarly, based on the system architecture, the scale-up and scale-out network
interconnects can be part of the same or different networks. The topology of the
inter-node and intra-node networks (fat-tree, dragonfly) is irrelevant to the
communication discussion exhibited in this work.

Distributed memory parallelism with the SPMD (Single Program Multiple Data)
programming style is the key focus for the communication operations addressed
in this work. This model for parallelism demonstrates the following
characteristics: (1) a set of tasks associated with a distributed application,
(2) tasks parts of a distributed application use their local memory during
computation, and (3) can reside on the same compute node or distributed across
multiple nodes. Tasks exchange data during communication. The communication
operations can be synchronous (\texttt{send} and \texttt{receive}) or
asynchronous (\texttt{put} and \texttt{get}) based on the distributed
programming model and the runtime systems used in the application.
Fig.~\ref{fig:hetero-systems} shows a simple
SPMD-style application with 8 tasks distributed across different compute nodes
in the system.

\subsection{Data and Control Path}\label{src:dp-cp}
Communication operations in GPU-attached memory regions discussed in this work
typically comprise \textit{data paths} and \textit{control paths}. Data paths
refer to those operations that involve moving data between the host-attached and
the GPU-attached memory regions. These data movement operations can occur within
the same compute node or between different compute nodes across a high-speed
network. Control paths correspond to coordination operations that occur between
(1) the application running on the host process, (2) the application compute
kernels running on the GPU, and (4) the NIC.

This analysis focuses on the semantics describing the control path of the data
movement operations when the memory buffer in the operation resides on the
accelerator memory domain (specifically GPU-attached memory). This work
classifies the various communication schemes used to perform the data movement
operations involving the GPU-attached memory buffers in the distributed
application. Section~\ref{src:intra-comms} and Section~\ref{src:inter-comms}
provide a background on the intra-node and inter-node communication performed on
the scale-up and scale-out interconnect, respectively. It shows the various
steps involved in performing the data movement operation.

\subsection{Communication Across Scale-up Interconnect}\label{src:intra-comms}
A scale-up interconnect connects the various components within a compute node.
It usually supports an extensive network bandwidth. The scale of the scale-up
network can vary from a small number of components, such as 4/8 GPUs (as in a
fat-node architecture seems in most HPC systems), to 256 GPUs (as in Nvidia POD
architecture) based on the type of the node architecture. Components within a
single compute node connected using the scale-up network are usually of the same
shared memory domain with or without cache coherency. This allows the
communication runtime and programming models supporting the required
communication operations to implement the required data movement \textit{load}
and \textit{store} operations.

\subsection{Communication Across Scale-Out Interconnect}\label{src:inter-comms}
Most scale-out interconnects connect the various compute nodes on the system.
These scale-out interconnects are usually linked to the compute units in each
compute node through a PCIe interface, and there can be multiple network
endpoints linked per compute node called \textit{multi-rail} systems. For
example, Fig.~\ref{fig:hetero-systems} represents two network endpoints per
compute node. Scale-out interconnects connecting the various compute nodes
usually support less network bandwidth when compared to the scale-up
interconnect due to the scale of the network that can sometimes extend beyond
8K nodes, as seen in many leadership supercomputers\cite{frontier1,perlmutter,aurora,alps}.

\begin{figure}[!ht]
  \centering
  \includegraphics[width=\linewidth]{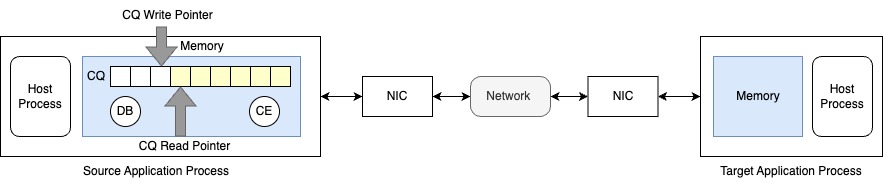}
  \caption{Representing a simple data movement from a source to target
  process in a scale-out network.}
  \label{fig:scale-out-comms-path}
\end{figure}

Unlike the communication operation supported in scale-up networks using load
and store operations, communication in scale-out networks requires complicated
packet transfer. Fig.~\ref{fig:scale-out-comms-path} represents the steps in
performing the data movement operation in a scale-out network. As shown in
Fig.~\ref{fig:scale-out-comms-path}, the data is moved from the source to the
target process. The source and target process are in a separate compute node,
with the data expected to be transferred through the scale-out network.

In this example, the data is initially available in the source process. The
source process that runs on a host process (like a CPU thread or core) initiates
the communication operation. The communication operation in this context can be
an RDMA operation, where the NIC associated with the host process retrieves the
data from the CPU-attached memory without the host or the operating system's
involvement and moves the data to the target process memory.

As a first step in this data movement operation, the host associated with the
source process creates a network command and enqueues it into a command queue
(CQ). The network command is a structure that contains the metadata for the NIC
to execute the required communication operation without the host or OS
involvement. The metadata in the network command could include properties such
as the target process information and the location and size of the memory buffer
on the source process that is expected to be transferred.

CQ is a memory location on the source process monitored by both the host and the
NIC using write and read pointers. The host process uses a write pointer to
identify the location in the CQ and enqueue the network command.

Once the network command is enqueued into the CQ, the host process triggers the
doorbell (DB) in the network to enable the NIC attached to the source to execute
the communication operation. Triggering the DB moves the read pointer in the CQ
for the NIC to identify the newly enqueued network command. Similarly, it moves
the write pointer to enable enqueuing new network commands.

Once the data is delivered, an acknowledgment for the delivered communication
operation is returned to the source process as an \texttt{ack}. The received ack
updates a completion event (CE) in the source process. CT can be used to track
the message completions. CE can be a simple monotonously increasing counter
update or a full-completion event enqueued into a completion queue.

Most scale-out interconnects provide support for network transfers as described
above. For brevity, the target side delivery of the message and the connection
establishment between the source and the target process are not described in
this work. With the intra-node and inter-node communication executed on scale-up
and scale-out networks introduced, the rest of this work classifies the various
available communication schemes for performing the data movement operations
involving the GPU-attached memory buffers. This includes a description of the
data path involved in the operation and the owner (CPU vs.\ GPU) of the control
path of the communication operation. Section~\ref{src:comms-baseline} and
Section~\ref{src:schemes} provide a description of these various communication
schemes.

%% file: src/content/schemes.tex
\section{Basic Communication Schemes}\label{src:comms-baseline}
This section describes the basic communication schemes supporting the
communication operations involving GPU-attached memory buffers across
different programming models and runtime systems. The control path
of the operation in the communication schemes discussed in this section is
managed by the host process (like a CPU thread or a core).

\subsection{Non GPU-Aware Communication}\label{src:non-gpu-aware}
\begin{figure}[!ht]
  \centering
  \includegraphics[width=\linewidth]{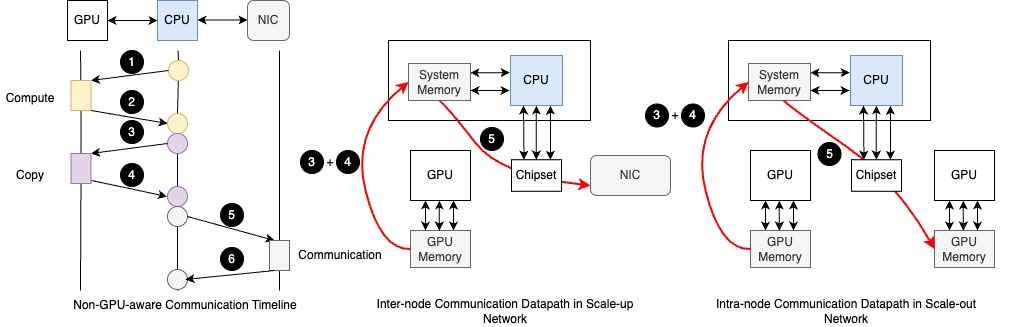}
  \caption{Representing non-GPU-aware communication scheme.}
  \label{fig:non-gpu-aware}
\end{figure}

A non-GPU-aware communication scheme involves a data movement operation with
GPU-attached memory buffers. This scheme requires the host process to
orchestrate the communication and manage the operation's control path and
datapath from the GPU-attached memory. It requires the data from the
GPU-attached memory to be staged into the CPU-attached system memory managed by
the host. Fig.~\ref{fig:non-gpu-aware} represents the various steps involved in
the non-GPU-aware communication scheme.

From the application perspective, as represented by the timeline chart in
Fig.~\ref{fig:non-gpu-aware}, a non-GPU-aware communication scheme has 6
(\begin{tiny}\circled[text=white,fill=black,draw=black]{1}\end{tiny} -
\begin{tiny}\circled[text=white,fill=black,draw=black]{6}\end{tiny}) steps. This
does not include initiating the GPU-attached buffers before executing the
compute operation.
Step \begin{tiny}\circled[text=white,fill=black,draw=black]{1}\end{tiny}
represents the host process launching the compute kernel to be executed on the
GPU, and step \begin{tiny}\circled[text=white,fill=black,draw=black]{2}\end{tiny}
represents the host waiting for the completion of the previously initiated
compute kernel on the GPU. While the compute kernel is being executed, the host
process is shown to wait for its completion. The host is
blocked from performing other operations between
\begin{tiny}\circled[text=white,fill=black,draw=black]{1}\end{tiny} and
\begin{tiny}\circled[text=white,fill=black,draw=black]{2}\end{tiny} while the
compute kernel is being executed.

Once the compute kernel is completed, the data associated with the compute
kernel will still reside in the GPU-attached memory. Before initiating a
non-GPU-aware communication operation, the host must stage the data from the
GPU-attached memory to the host-managed CPU-attached memory. This is performed
using a copy operation as shown in
\begin{tiny}\circled[text=white,fill=black,draw=black]{3}\end{tiny} and
\begin{tiny}\circled[text=white,fill=black,draw=black]{4}\end{tiny}. The host
process waits on the completion of the copy operation and is blocked from
performing other operations. With the data now staged into the CPU-attached
memory, the host process initiates the communication operation in step
\begin{tiny}\circled[text=white,fill=black,draw=black]{5}\end{tiny} and
waits for its completion in step
\begin{tiny}\circled[text=white,fill=black,draw=black]{6}\end{tiny}.
The core semantics of the non-GPU-aware communication scheme is the following:
\begin{enumerate}
    \item The host process involved in the communication initiates the data
    movement operation.
    \item It is not possible to directly move the data from the GPU-attached
    memory. Instead, the host process orchestrating the communication operation
    is expected to stage the source buffer from the GPU-attached memory into
    CPU-attached system memory before initiating the data movement operation. And,
    \item The staging of the source buffer into the CPU-attached system memory
    (\begin{tiny}\circled[text=white,fill=black,draw=black]{3}\end{tiny} and
    \begin{tiny}\circled[text=white,fill=black,draw=black]{4}\end{tiny})
    is required for both intra-node and inter-node operation on the scale-up and
    scale-out networks, respectively.
\end{enumerate}

While the non-GPU-aware communication scheme introduces the GPU-based
communication operation involving GPU-attached memory buffers, the baseline
GPU-aware communication is the state-of-the-art implementation supported across
different programming models and runtime systems. The baseline GPU-aware
communication is described in Section~\ref{src:gpu-awareness}.

\subsection{Baseline GPU-Aware Communication}\label{src:gpu-awareness}

\begin{figure}[!ht]
  \centering
  \includegraphics[width=\linewidth]{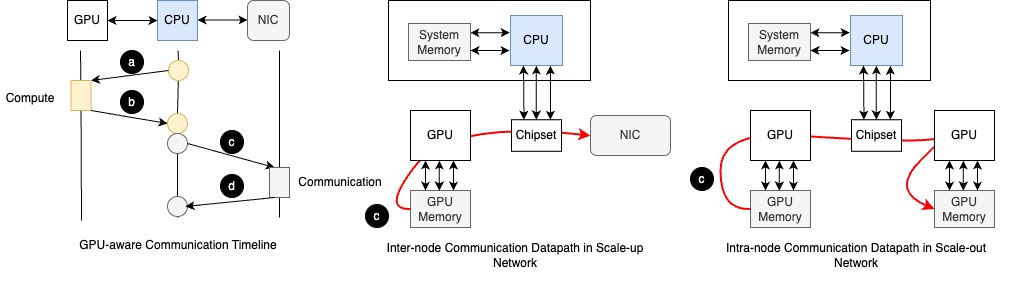}
  \caption{Representing GPU-aware communication scheme.}
  \label{fig:gpu-aware}
\end{figure}

A GPU-aware communication scheme~\cite{mvapich2-gpu-comp-science-journal,
sreeram-internode-ICPP,sreeram-ipdps12-cuda-ipc,gpu-aware-non-contig-cluster12,
gpu-shmem-osm-workshop-2015} extends the non-GPU-aware communication scheme
described in section~\ref{src:non-gpu-aware}. In this communication scheme,
staging the GPU-attached memory buffer into the host-managed system memory is
not needed. Fig.~\ref{fig:gpu-aware} represents the various steps involved in the
GPU-aware communication scheme.

As shown in Fig.~\ref{fig:gpu-aware}, the need for staging the source buffer
from the GPU-attached memory to the CPU-attached memory is eliminated.
Irrespective of the intra-node and inter-node communication operation, GPU-aware
communication from the application perspective allows the data to be directly
transferred to the target process using a zero-copy communication model. Steps
\begin{tiny}\circled[text=white,fill=black,draw=black]{a}\end{tiny} and
\begin{tiny}\circled[text=white,fill=black,draw=black]{b}\end{tiny} in
Fig.~\ref{fig:gpu-aware} represent the host process launching the compute
kernel and waiting for completion. With the compute kernel completed, data
associated with the compute kernel that residess on the GPU-attached memory is directly
used for the communication operation. This is shown in
step \begin{tiny}\circled[text=white,fill=black,draw=black]{c}\end{tiny} and step
\begin{tiny}\circled[text=white,fill=black,draw=black]{d}\end{tiny}.
The core semantics of the GPU-aware communication scheme is the following:
\begin{enumerate}
    \item The host process involved in the communication initiates the data
    movement operation.
    \item The datapath in the communication scheme does not necessitate staging
    the GPU-attached buffer into the system-managed CPU-attached memory. Hence,
    the data associated with the compute kernel resides in the GPU-attached
    memory and is directly used for communication. And,
    \item The source buffer can be communicated without staging into the
    host-managed system memory
    (\begin{tiny}\circled[text=white,fill=black,draw=black]{c}\end{tiny} and
    \begin{tiny}\circled[text=white,fill=black,draw=black]{d}\end{tiny})
    for both intra-node and inter-node operation on the scale-up and scale-out
    networks, respectively.
\end{enumerate}

The basic non-GPU-aware and GPU-aware communication scheme described in sections
~\ref{src:non-gpu-aware} and ~\ref{src:gpu-awareness} represents the basic
communication schemes where the host process orchestrates the communication
operation, and the control path of the operation is managed by the host process
involved in the communication operation. There are multiple runtime libraries
and programming models that support these communication schemes. The GPU-centric
communication schemes discussed in section~\ref{src:schemes} represent the
advanced communication schemes where the
control path of the communication operation is moved from the CPU to the GPU.

\section{GPU-Centric Communication Schemes}\label{src:schemes}
As mentioned in Section~\ref{src:comms-baseline}, a GPU-centric communication
scheme involves the GPU managing the control path of the communication operation.
This section describes three different GPU-centric communication schemes.

\subsection{Stream Triggered Communication}\label{src:st}
Stream Triggered (ST) communication
scheme~\cite{naveen-st-p2p,naveen-st-arma,mpix-stream-anl,euro-mpi-joseph}
enables a GPU-aware application to offload the control paths of the
communication operations to the underlying implementation and hardware
components. This specifically includes the \textit{GPU Stream}.

\subsubsection{GPU Streams}
A GPU stream~\cite{cuda-stream} is a queue of device operations. GPU compute
kernel concurrency is achieved by creating multiple concurrent streams.
Operations issued on a stream typically run asynchronously with respect
to the CPU and operations enqueued in other GPU streams. Operations in a given
stream are guaranteed to be executed in FIFO order. In this work, the GPU
component that provides these execution guarantees to schedule and control the
execution of the enqueued operation is referred to as the \textit{GPU Stream
Execution Controller} (GPU SEC). Depending on the GPU vendor, GPU SEC can be a
software, hardware, or kernel component associated with the GPU. ST
scheme enables offloading the control path of the operation involving
the GPU-attached memory buffers into the GPU SEC.

\subsubsection{ST Description}
A parallel application using the ST scheme continues to manage compute kernels
on the GPU via existing mechanisms. In addition, the ST scheme allows an
application process running on the CPU to define a set of ST communication
operations. These communication operations can be scheduled for execution at a
later point in time. More importantly, in addition to offering a deferred
execution model, ST enables the GPU to get closely involved in the control
paths of the communication operations.

\begin{figure}[ht]
  \includegraphics[width=0.7\linewidth]{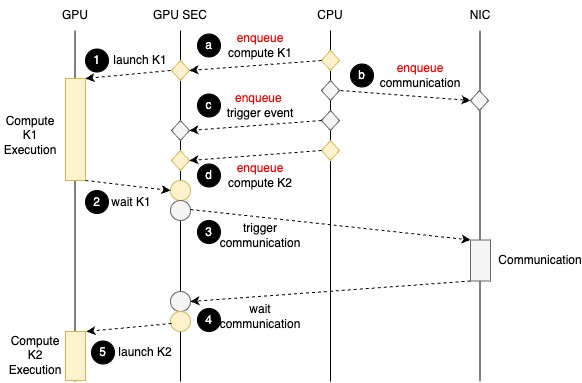}
  \caption{Representing a GPU-aware application using ST.}
  \label{fig:st-execution}
\end{figure}

Fig.~\ref{fig:st-execution} illustrates a sequence of events involved in a
parallel application using ST inter-process communication and synchronization
operations. An application process running on the CPU enqueues
\begin{tiny}\circled[text=white,fill=black,draw=black]{a}\end{tiny} GPU
kernel K1 to the GPU stream,
\begin{tiny}\circled[text=white,fill=black,draw=black]{b}\end{tiny} triggered
ST-based communication operations to the NIC,
\begin{tiny}\circled[text=white,fill=black,draw=black]{c}\end{tiny} the
corresponding trigger event to the GPU stream, and
\begin{tiny}\circled[text=white,fill=black,draw=black]{d}\end{tiny} GPU kernel
K2 to the same stream. The CPU returns immediately after the operations are
enqueued and is not blocked on the completion of the enqueued operations.
It is the GPU SEC responsibility to
\begin{tiny}\circled[text=white,fill=black,draw=black]{1}\end{tiny} launch K1
and \begin{tiny}\circled[text=white,fill=black,draw=black]{2}\end{tiny} wait for
its completion. Once K1 completes,
\begin{tiny}\circled[text=white,fill=black,draw=black]{3}\end{tiny} GPU
SEC triggers the execution of previously enqueued communication operations and
\begin{tiny}\circled[text=white,fill=black,draw=black]{4}\end{tiny} waits for
these operations to finish. Next,
\begin{tiny}\circled[text=white,fill=black,draw=black]{5}\end{tiny} the GPU
SEC launches K2.

The key semantics of the ST scheme includes the following:
\begin{enumerate}
    \item The CPU offloads the control path of the communication operation to
    the GPU SEC.
    \item While the CPU initiates the communication operation, the GPU SEC
    triggers the execution of the previously enqueued communication operation. And,
    \item Communication involving the GPU-attached memory buffers are executed
    at GPU kernel boundaries.
\end{enumerate}

With the ST communication scheme, an application process running on the CPU
enqueues operations to the NIC command queue (as described in
Section~\ref{src:inter-comms}) and the GPU stream, but does not get
directly involved in the control paths of communication operations,
subsequent kernel launch, and tear-down operations.
The CPU does not directly wait for communication operations to complete.
The GPU manages the control paths and eliminates potential synchronization
points in the application.

\subsection{Kernel Triggered Communication}\label{src:kt}

Kernel Triggered (KT) communication scheme is an extension to the ST scheme.
Offloading the communication control path from the CPU to GPU SEC in ST allows
the CPU to exit the communication operation without involved in the
communication control path. While offloading the communication control path to
the GPU is useful, communication in ST still happens at compute kernel
boundaries. KT allows the communication to be performed from within the
compute kernel. KT offloads the communication control path to the GPU thread or
thread-block based on the granularity of the communication operation.

\begin{figure}[ht]
  \includegraphics[width=0.9\linewidth]{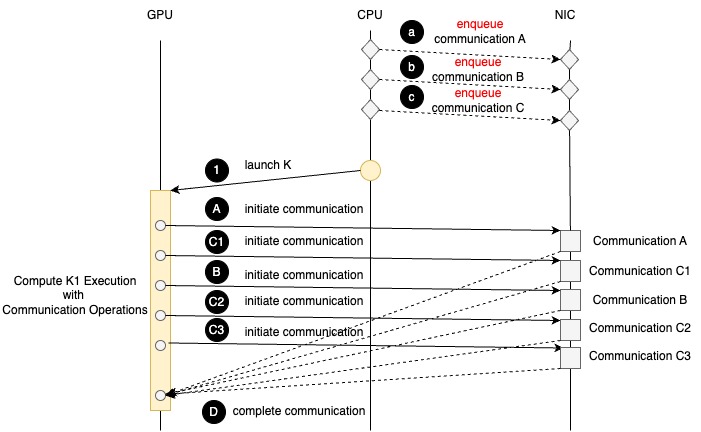}
  \caption{Representing various events in a GPU-aware application using KT.}
  \label{fig:kt-execution}
\end{figure}

Fig.~\ref{fig:kt-execution} shows the various steps involved in the KT
communication scheme. This example shows three communication operations
(\begin{tiny}\circled[text=white,fill=black,draw=black]{a}\end{tiny},
\begin{tiny}\circled[text=white,fill=black,draw=black]{b}\end{tiny}, and
\begin{tiny}\circled[text=white,fill=black,draw=black]{c}\end{tiny})
getting enqueued into the NIC. The host process associated with the application
enqueues the communication operation along with enqueuing the compute kernel
\textit{K}.
Once both the communication and compute operations are enqueued, the host
process returns immediately. It is good to note that the communication
operations are enqueued before the compute kernel. With the compute kernel
executed, previously enqueued communication operations can be triggered within
the compute kernel. Steps
\begin{tiny}\circled[text=white,fill=black,draw=black]{A}\end{tiny} and
\begin{tiny}\circled[text=white,fill=black,draw=black]{B}\end{tiny} show the
GPU triggering the execution of operation
\begin{tiny}\circled[text=white,fill=black,draw=black]{a}\end{tiny} and
\begin{tiny}\circled[text=white,fill=black,draw=black]{b}\end{tiny}
respectively. Similarly, communication operation
\begin{tiny}\circled[text=white,fill=black,draw=black]{c}\end{tiny} can be
considered as a persistent operation getting triggered for execution multiple
times
(\begin{tiny}\circled[text=white,fill=black,draw=black]{C1}\end{tiny},
\begin{tiny}\circled[text=white,fill=black,draw=black]{C2}\end{tiny}, and
\begin{tiny}\circled[text=white,fill=black,draw=black]{C3}\end{tiny}) from
within the compute kernel.
The key semantics of the KT scheme includes the following:
\begin{enumerate}
    \item The CPU offloads the control path of the communication operation to
    the GPU.
    \item While the CPU initiates the communication operation, the GPU triggers
    executing the previously enqueued communication operation.
    \item Communication involving the GPU-attached memory buffers is executed
    within a GPU kernel and not across GPU kernel boundaries. And,
    \item KT supports managing persistent communication operation.
\end{enumerate}

\subsection{Kernel Initiated Communication}\label{src:ki}

Kernel Initiated (KI) communication enables the GPU to initiate and trigger the
execution of the communication operation associated with a GPU-attached memory
buffers. Similar to the KT scheme, there are different granularities in
initiating the communication operation like the GPU thread-level and GPU
thread-block level operations. For this section, the communication granularity
is not considered.

\subsubsection{KI Description}
KI communication
scheme~\cite{ipdpsw-nvshmem-2020,gio-amd,gpu-verbs,gpu-centric-ib-hipc-2017,intel-shmem,offload-comms-control}
involves the host process part of the application to initiate the compute kernel
and exit from any further processing. With the GPU executing the enqueued
compute kernel, it can initiate and execute communication operations from within
the compute kernels. In this scheme, the GPU thread/thread-block initiates the
communication operation by preparing the network packets necessary for
performing the inter-node operation or creating an inter-process shared
memory-based communication across the GPU/CPU components in the compute node.
The KI scheme provides complete GPU autonomy when performing communication
operations.

\begin{figure}[ht]
  \includegraphics[width=0.9\linewidth]{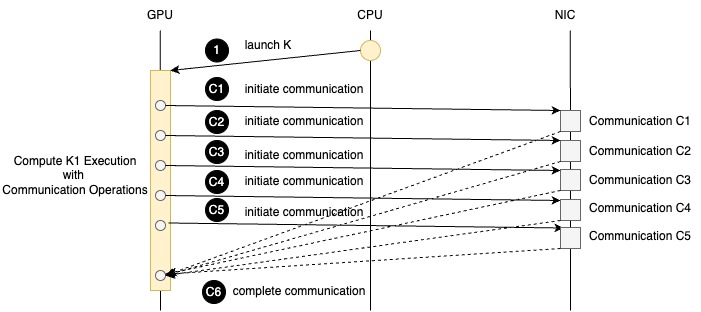}
  \caption{Representing various events in a GPU-aware application using KI.}
  \label{fig:ki-execution}
\end{figure}

Fig.~\ref{fig:ki-execution} shows the various steps involved in KI communication
scheme. As shown in Fig.~\ref{fig:ki-execution},
\begin{tiny}\circled[text=white,fill=black,draw=black]{1}\end{tiny}
shows the host process involved in the application launches the compute kernel
(\textit{K}) into the GPU execution stream and returns immediately. Within the
compute kernel, the GPU thread or thread-blocks can initiate and execute the
communication operations
(\begin{tiny}\circled[text=white,fill=black,draw=black]{C1}\end{tiny} -
\begin{tiny}\circled[text=white,fill=black,draw=black]{C5}\end{tiny})
without involving the CPU in the communication control path. GPU manages both
the communication execution and synchronization, determining the completion of
the executed operations. In this example, a course-grain completion is shown as
step \begin{tiny}\circled[text=white,fill=black,draw=black]{C6}\end{tiny}.
The core semantics of the KI scheme involves the following:
\begin{enumerate}
    \item KI communication operations are initiated and executed by the GPU.
    \item The GPU manages the communication control path and orchestrates data
    movement without CPU assistance.
    \item Communication operations initiated by the GPU are immediately executed
    and are not deferred execution operations. And,
    \item Communication is performed within the compute kernel. It is not
    necessary to tear-down the compute kernels to perform the communication
    operation.
\end{enumerate}

\subsection{Comparing Communication Schemes}\label{src:compare}
With various basic and GPU-centric communication schemes exibited in
section~\ref{src:comms-baseline} and section~\ref{src:schemes}, this section
compares the various communication traits across these different communication
schemes. Table~\ref{tab:comms-scheme} provides the result of the analysis.

\begin{table*}[ht]
\centering
\caption{Table comparing different communication schemes.}
\label{tab:comms-scheme}
\begin{tabular}{|l|l|l|l|l|l|}
\hline
 &
  \multicolumn{1}{c|}{\textbf{NGA}} &
  \multicolumn{1}{c|}{\textbf{GA}} &
  \multicolumn{1}{c|}{\textbf{ST}} &
  \multicolumn{1}{c|}{\textbf{KT}} &
  \multicolumn{1}{c|}{\textbf{KI}} \\ \hline
\multirow{2}{*}{Orchestrator} &
  \multirow{2}{*}{CPU} &
  \multirow{2}{*}{CPU} &
  \multirow{2}{*}{GPU SEC} &
  \multirow{2}{*}{GPU} &
  \multirow{2}{*}{GPU} \\
                      &     &     &         &       &         \\ \hline
Initiator             & CPU & CPU & CPU     & CPU   & GPU     \\ \hline
Executor              & CPU & CPU & GPU SEC & GPU   & GPU     \\ \hline
Execution point &
  \begin{tabular}[c]{@{}l@{}}Kernel\\ Boundary\end{tabular} &
  \begin{tabular}[c]{@{}l@{}}Kernel\\ Boundary\end{tabular} &
  \begin{tabular}[c]{@{}l@{}}Kernel\\ Boundary\end{tabular} &
  \begin{tabular}[c]{@{}l@{}}Within\\ Kernel\end{tabular} &
  \begin{tabular}[c]{@{}l@{}}Within\\ Kernel\end{tabular} \\ \hline
Direct copy             & No  & Yes & Yes     & Yes   & Yes     \\ \hline
Communication Pattern & N/A & N/A & Fixed   & Fixed & Dynamic \\ \hline
\end{tabular}
\end{table*}

As shown in Table~\ref{tab:comms-scheme}, the different exibited communication
schemes are compared against the different communication traits such as the
component responsible for initiating, executing and orchestrating the
communication operation, the execution point in the application, need for staging
the data, and communication pattern suitable for the schemes. With the different
communication schemes introduced in section~\ref{src:schemes}, the various
factors impacting the implementation of the different communication schemes and
the communication patterns addressed by the communication schemes are discussed
in section~\ref{src:impact}.

%% file: src/content/implement.tex
\section{Factors Impacting GPU Communication Schemes}\label{src:impact}
This section analyzes the contemporary heterogeneous system architecture
landscape and the most common communication pattern involving GPU-attached
memory buffers executed on these systems. The goal is to expose the criticality
of the GPUs in modern supercomputing systems and the different types of network
interconnects used to link the various components in the compute nodes that
could impact the performance and functionality of the required data movement
operations.

\subsection{Heterogeneous System Landscape}\label{src:systems}
The major components of a heterogeneous HPC compute node as introduced in
section~\ref{src:bground} include (a) a host processor (like a central
processing unit (CPU)), (b) an accelerator (like a graphics processing unit
(GPU)), and (c) a network interconnect (NIC) connecting the various components
of the compute node and across the network connecting different compute nodes in
the system. Some prominent heterogeneous node architectures from the
Top500~\cite{top500} list are used for this analysis.

\begin{figure}[!ht]
  \centering
  \includegraphics[width=\linewidth]{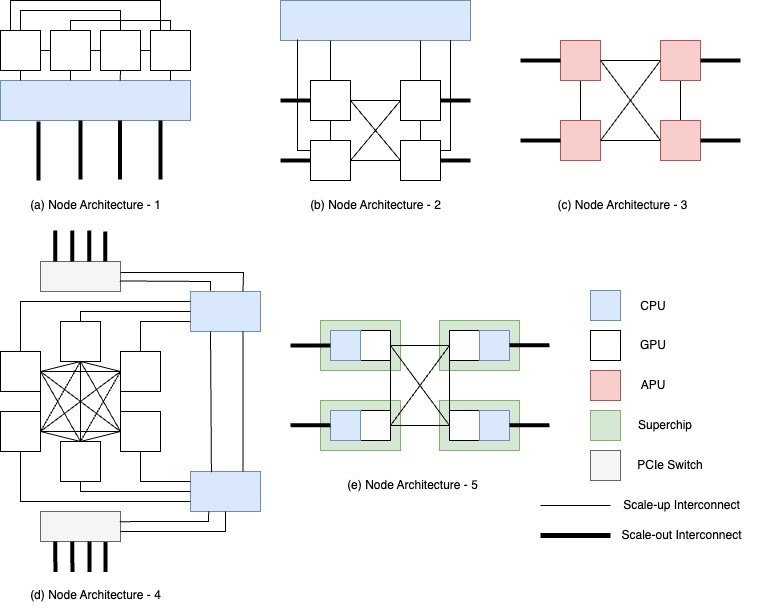}
  \caption{Modern heterogeneous system node architectures with network links.}
  \label{fig:hetero-nodes}
\end{figure}

Fig.~\ref{fig:hetero-nodes} represents the different node architectures used for
this discussion. The network interconnect usage in these architectures is
tightly linked to support the communication schemes discussed in this work. In
some modern architectures, the discrete CPUs and GPUs are replaced with
specialized processors like an APU or a \textit{superchip}. In brief, an APU is
a CPU built-in with a GPU on a single die. While a superchip is similar in
design with an APU, the CPU and GPU are interconnected using an high-speed
network interconnect in this design.

Examples of different components include Intel Xeon~\cite{intel-xeon},
AMD Milan~\cite{amd-milan}, and AMD Genova~\cite{amd-genova} for CPUs,
AMD MI250X, Nvidia A100~\cite{a100}, and Intel Max Series GPUs
code-named \textit{Ponte Vecchio}~\cite{pvc} for GPUs, AMD MI300A~\cite{mi300a}
for APUs, Nvidia Grace-Hopper~\cite{gh200} for superchips, and Nvidia
Infiniband~\cite{infiniband}, HPE Slingshot~\cite{slingshot}, AMD Infinity
Fabric~\cite{infinity-fabric}, Nvidia NVlink~\cite{nvlink}, and
Ethernet for network interconnects. Fig.~\ref{fig:hetero-nodes}
provides a high-level representation of five heterogeneous node architectures.
For brevity, vendor details and compute capabilities of each component in the
architecture are not mentioned in Fig.~\ref{fig:hetero-nodes}.


The following are the core observations across these different architectures and
their impact in supporting the different GPU-centric communication schemes used
for the data movement operations in these systems.

\subsubsection{\textbf{Number of GPUs}}
Most node architectures exhibited in Fig.~\ref{fig:hetero-nodes} support
multiple accelerators (GPUs or APUs) per compute node. With respect to
GPU-centric communication, this is critical if the data movement operation
within a node is managed by a high-speed and high-bandwidth scale-up NIC like
NVlink and Infinity Fabric. The programming models and runtime systems
supporting the GPU-centric communication schemes are expected to support an
independent communication middleware in the software stack exploiting the
shared memory domain supported by the scale-up NIC.

\subsubsection{\textbf{CPU to GPU Ratio}}
The ratio of the GPUs to the CPUs across most of the architectures is
high ($4:1$ or $6:2$), Except \texttt{Node Architecture-3} and
\texttt{Node Architecture-5} which represents $1:1$ GPU:CPU ratio due to the
inherent APU and superchip design. While it is ideal to offload the GPU-centric
communication operations to the underlying NIC, supporting these ideal
implementation designs across all system architectures is not possible. In
NICs where the offload capability is not supported, one or more processes
(core or thread) from the host are used as an asynchronous progress thread to
implement the data movement and synchronization operations. CPU to GPU ratio
determines the resources available to support such emulation design for
implementing GPU-centric communication schemes.

\subsubsection{\textbf{Number of NICs}}
There are two different networks in the system: a scale-up network connecting
the various components within the node and a scale-out network connecting the
node with other nodes in the system. Effectively utilizing the available
networks is critical for performing the data movement operations involving the
GPU-attached memory buffers.

\subsubsection{\textbf{Scale-out NIC Connection}}
The compute node component on which the scale-out NIC is attached is not
uniform. In some architectures, the NIC is directly attached to the GPU
(\texttt{Node Architecture-2} and \texttt{Node Architecture-3}), while there
are architectures where the NIC is attached to the CPU
(\texttt{Node Architecture-1} and \texttt{Node Architecture-5}) or a PCIe
switch (\texttt{Node Architecture-4}). The latency of the small payload
communication operations is usually influenced by this design.

\subsubsection{\textbf{GPU to NIC Ratio}}
Most node architectures support a $1:1$ GPU to NIC ratio. This allows the user
libraries to support an application process per GPU. When more than one NIC is
available per GPU, the user libraries are expected to support multiple NICs per
process or force applications to adopt a multiple-process per GPU job
distribution. The GPU-to-NIC ratio determines the type of process configuration
associated with the GPU-centric communication schemes.

\subsubsection{\textbf{Impact of PCIe Switching}}
While most scale-up NICs are directly connected to a PCIe root complex,
specifically in the \texttt{Node Architecture-4}, the NICs are connected to a
PCIe switch. This design decision can impact the various features supported by
the NIC, such as peer mapping the NIC resources in the GPU memory domain, and,
in turn, impact the GPU-centric communication schemes. For example, features
supporting network address translation impacting the RDMA communication and
network caching of memory registered with the NIC to perform the RDMA operation
can be impacted by this design.

Overall, it is essential to understand the criticality of the compute node
architecture determines the host, accelerator, and NICs that link all the
components within the compute node. The compute node architecture determines
the proposed GPU-centric communication scheme implementation options. In this
section, we briefly discussed the various compute node architecture features
that could impact the performance and functionality of the GPU-centric
communication schemes. Section~\ref{src:implement} introduces the various GPU and
NIC capabilities required to implement the exhibited GPU-centric communication
schemes and section~\ref{src:apps} classifies the various
communication patterns the different communication schemes address.

\subsection{Implementation Requirements}\label{src:implement}

This section discusses the essential features required for implementing the
communication schemes described in section ~\ref{src:schemes}. While multiple
factors could impact the performance and functionality of the communication
operations using GPU-attached memory buffers, features described in this
section can be considered critical. It is important to note that some of the
terminologies used in the section are specific to a particular GPU vendor (
Nvidia), but most known GPU vendors (specifically AMD) have support for all the
below-mentioned features.

\subsubsection{\textbf{GPUDirect Peer to Peer}}\label{src:peer-to-peer}
Peer-to-peer communication allows the communication runtime (like Cray MPI) to
implement zero-copy for many, but not necessarily all, data transfers across
PEs sharing the same shared memory domain (intra-node communication). This
feature is also referred to as GPUDirect P2P. GPUDirect Peer to Peer is
supported natively by most known GPU vendors in their device drivers. This
feature enables GPU-to-GPU copies, loads, and stores directly over the memory
fabric (PCIe, NVLink, Infinity Fabric).

Multiple middleware libraries enable GPUDirect P2P. Inter-Process Communication
(IPC) is a capability supported by both CUDA~\cite{cuda} and HIP~\cite{hip}
runtimes that allows GPUDirect P2P for on-node data transfers. For fine-grained
communication control, AMD offers Heterogenous System Architecture (HSA)~\cite{
hsa}, a low-level library for performing IPC.

\subsubsection{\textbf{GPUDirect RDMA}}\label{src:gdrdma}
GPUDirect RDMA is a technology that allows a direct path for data exchange
between the GPU and a third-party peer device (like scale-out interconnect)
using standard features of PCI Express. GPUDirect RDMA is used to perform
zero-copy transfers of the GPU-attached memory buffers, when possible, from the
GPU to the NIC without staging through a CPU-attached memory buffer. Most known
device vendors provide support for GPUDirect RDMA and the scale-out
interconnect attached to the GPU can perform data transfers on the GPU-attached
memory buffers without staging through a host CPU-attached memory buffer.

Other intra-node memory copy libraries like GDRCopy~\cite{gdrcopy} for Nvidia
GPU devices are also capable of using the GPUDirect RDMA to perform effective
data transfers between the GPU and the CPU. GDRCopy creates a CPU mapping of
the GPU memory and uses it to perform low-latency data transfers between the
GPU and CPU.

GPUDirect RDMA is a critical feature required for performing as many data transfers as possible across different compute nodes and within the same compute node.

\subsubsection{\textbf{GPUDirect Async}}\label{src:dgasync}
A common semantic for initiating a network data transfer involves initiating
the communication to enqueue a network command entry into a predefined queue
(called a command queue) and notifying the NIC about the enqueue operation by
ringing a doorbell. This can trigger the NIC dequeue and execute the operation
defined by the process in the network command entry.

GPUDirect Async is a new feature supported across different devices to allow
the GPU to directly trigger and poll for completion of the communication
operations queued to a command queue managed by the NIC. This feature depends
on the ability to create GPU peer mappings of the NIC BAR space, which allows
the NIC MMIO registers to be mapped into the GPU memory space so the GPU
threads (for KT and KI) or the GPU SEC (for ST) can control the communication
operations.

GPUDirect Async is a critical feature for supporting data transfers across
different compute nodes using advanced GPU-centric communication schemes like
ST, KT, and KI. HPE Slingshot NIC and Nvidia Infiniband can support the
GPUDirect Async feature if the underlying GPUs provide it.

\subsubsection{\textbf{Triggered Communication Operations}}\label{src:tops}
Triggered communication operations are network data transfer operations with
deferred execution semantics. While a traditional data transfer involves the
NIC immediately executing the operation as soon as the process enqueuing the
data transfer rings the doorbell, notifying the enqueued network command entry,
a triggered operation defers the execution of the operation until an associated
condition to the network command entry is satisfied.

The conditions associated with a triggered operation include a network counter
matching a posted trigger threshold. When the counter reaches the expected
threshold value determined by the application, the associated data movement
operation is executed.

\subsubsection{\textbf{Network Descriptor Templating}}\label{src:template}
As briefly introduced in section~\ref{src:inter-comms}, posting a network
communication operation involves the source process generating the data
transfer to enqueue a network command entry into a command queue monitored by
the NIC. The placement of the command queue and the creator of the command
entry are critical to enable the kernel-initiated communication scheme to be
supported. Since the GPU thread or the thread-block is involved in generating
the command entry, the cost of this operation is heavy as the operation is
sequential.

Command entry templating allows the source process to maintain a predefined
descriptor that can be filled quickly and then copied into the command queue
relatively cheaply. This allows for an efficient implementation of the
kernel-initiated communication scheme. Scale-out interconnects like HPE
Slingshot NIC have different options to support this expected feature in
enabling kernel-initiated communication.

\subsubsection{\textbf{Network Interconnect On-Demand Paging}}\label{src:odp}
Network memory registration (MR) is a mechanism that allows an application to
describe a virtually contiguous memory location to the network adapter. The MR
process pins the memory pages to avoid getting swapped and maintain the
associated physical-to-virtual memory mapping. It is a key to executing RDMA
operations. It creates a \textit{key} (label) to the registered memory buffer
by setting specific permissions. The key details can be transferred to a remote
process in the application to perform remote RDMA operation.

While the MR process is critical for RDMA, it is usually considered a major
inhibitor to RDMA adoption where the programming model or runtime systems do
not naturally support the MR process in the exposed semantics.

Pinning the memory as part of the MR process can be critical for implementing
GPU-centric communication operations using GPU-attached memory buffers. It can
restrict the use of advanced GPU-attached memory types, such
as managed memory~\cite{managed-memory} and unified virtual memory~\cite{uvm}.
These advanced GPU memory types require swapping pages between the CPU and GPU
to allow efficient programmability. Registering these memory types for
GPU-centric communication is tricky.

On-Demand-Paging (ODP) is a technique that eases memory registration.
Applications do not need to pin down the underlying physical pages of the
address space and track the validity of the mappings. Modern NICs like
Inifiband and HPE Slingshot can support ODP efficiently and ease the
implementation of GPU-centric communication operations.

This section briefly discussed the critical GPU and NIC capabilities that are
key for efficiently implementing GPU-centric communication schemes exhibited in
section~\ref{src:schemes}. Section~\ref{src:user-libs} provides a list of
libraries and runtime systems supporting the GPU-centric communication schemes
using the GPU and NIC capabilities discussed in this section.

\subsection{GPU-centric Communication Libraries}\label{src:user-libs}
This section briefly discusses the various communication libraries and runtime
systems supporting different GPU-centric communication schemes. For brevity, this list does not include the various GPU-aware and non-GPU-aware communication libraries along with middlewares (like libfabric~\cite{libfabric}, ibverbs~\cite{ibverbs}, and UCX~\cite{ucx}) exposing the various GPU-centric communication semantics.

\begin{table*}[ht]
\centering
\caption{Table representing the different user-level communication libraries and
runtime systems supporting GPU-centric communication schemes.}
\label{tab:libs}
\begin{tabular}{|l|l|l|l|lll|lll|}
\hline
\multicolumn{1}{|c|}{\multirow{2}{*}{\textbf{Name}}} &
  \multicolumn{1}{c|}{\multirow{2}{*}{\textbf{ST}}} &
  \multicolumn{1}{c|}{\multirow{2}{*}{\textbf{KT}}} &
  \multicolumn{1}{c|}{\multirow{2}{*}{\textbf{KI}}} &
  \multicolumn{3}{c|}{\textbf{Operations}} &
  \multicolumn{3}{c|}{\textbf{Devices}} \\ \cline{5-10}
\multicolumn{1}{|c|}{} &
  \multicolumn{1}{c|}{} &
  \multicolumn{1}{c|}{} &
  \multicolumn{1}{c|}{} &
  \multicolumn{1}{c|}{\textbf{P2P}} &
  \multicolumn{1}{c|}{\textbf{RMA}} &
  \multicolumn{1}{c|}{\textbf{COLL}} &
  \multicolumn{1}{c|}{\textbf{AMD}} &
  \multicolumn{1}{c|}{\textbf{Nvidia}} &
  \multicolumn{1}{c|}{\textbf{Intel}} \\ \hline
NCCL &
  $\checkmark$ &
   &
   &
  \multicolumn{1}{l|}{$\checkmark$} &
  \multicolumn{1}{l|}{} &
  $\checkmark$ &
  \multicolumn{1}{l|}{} &
  \multicolumn{1}{l|}{$\checkmark$} &
   \\ \hline
RCCL &
  $\checkmark$ &
   &
   &
  \multicolumn{1}{l|}{$\checkmark$} &
  \multicolumn{1}{l|}{} &
  $\checkmark$ &
  \multicolumn{1}{l|}{$\checkmark$} &
  \multicolumn{1}{l|}{} &
   \\ \hline
OneCCL &
  $\checkmark$ &
   &
   &
  \multicolumn{1}{l|}{} &
  \multicolumn{1}{l|}{} &
  $\checkmark$ &
  \multicolumn{1}{l|}{} &
  \multicolumn{1}{l|}{} &
  $\checkmark$ \\ \hline
NVSHMEM &
  $\checkmark$ &
   &
  $\checkmark$ &
  \multicolumn{1}{l|}{} &
  \multicolumn{1}{l|}{$\checkmark$} &
   &
  \multicolumn{1}{l|}{} &
  \multicolumn{1}{l|}{$\checkmark$} &
   \\ \hline
ROC\_SHMEM &
  $\checkmark$ &
   &
  $\checkmark$ &
  \multicolumn{1}{l|}{} &
  \multicolumn{1}{l|}{$\checkmark$} &
   &
  \multicolumn{1}{l|}{$\checkmark$} &
  \multicolumn{1}{l|}{} &
   \\ \hline
Intel SHMEM &
   &
   &
  $\checkmark$ &
  \multicolumn{1}{l|}{} &
  \multicolumn{1}{l|}{$\checkmark$} &
   &
  \multicolumn{1}{l|}{} &
  \multicolumn{1}{l|}{} &
  $\checkmark$ \\ \hline
HPE Cray MPI &
  $\checkmark$ &
  $\checkmark$ &
   &
  \multicolumn{1}{l|}{$\checkmark$} &
  \multicolumn{1}{l|}{$\checkmark$} &
   &
  \multicolumn{1}{l|}{$\checkmark$} &
  \multicolumn{1}{l|}{$\checkmark$} &
   \\ \hline
MPICH &
  $\checkmark$ &
   &
   &
  \multicolumn{1}{l|}{$\checkmark$} &
  \multicolumn{1}{l|}{} &
   &
  \multicolumn{1}{l|}{$\checkmark$} &
  \multicolumn{1}{l|}{$\checkmark$} &
  $\checkmark$ \\ \hline
Libmp &
  $\checkmark$ &
  $\checkmark$ &
  $\checkmark$ &
  \multicolumn{1}{l|}{$\checkmark$} &
  \multicolumn{1}{l|}{$\checkmark$} &
   &
  \multicolumn{1}{l|}{} &
  \multicolumn{1}{l|}{$\checkmark$} &
   \\ \hline
\end{tabular}
\end{table*}

Table~\ref{tab:libs} provides a comprehensive list of various user-level
communication libraries and runtime systems supporting GPU-centric
communication schemes. As shown in Table~\ref{tab:libs}, the communication
libraries and runtime systems supporting the GPU-centric communication schemes
are supported across different GPU devices provided by vendors such as AMD,
Intel, and Nvidia. These libraries expose the GPU-centric communication schemes
through various operations supporting different communication protocols, such
as: (1) \textit{P2P} refers to the synchronous point-to-point communication
model exposing message passing operations like \textit{send} and
\textit{receive}, (2) \textit{RMA} refers to remote memory access-based
one-sided communication model supported by partitioned global address space
(PGAS) based style of programming supporting operations like \textit{puts} and
\textit{gets}, and collective communication operations like \textit{allreduce},
\textit{broadcast}, \textit{reduce-scatter}, and \textit{allgather} performed
by a group of processes.

Comparing and contrasting the suitability of different operations supporting
GPU-centric communication schemes in these libraries listed in
Table~\ref{tab:libs} and the availability of these libraries across different
network interconnects (like Nvidia Infiniband and HPE Slingshot) is beyond the
scope of this work. In brief, this section shows an early adaptability of the
various GPU-centric communication schemes exhibited in this work. Also, it
shows the various operations (P2P, RMA, and collectives) used to expose the
GPU-centric communication schemes to GPU-aware applications in different
domains such as HPC and ML. Section~\ref{src:apps} classifies the various
communication patterns addressed by the different communication schemes
introduced in section~\ref{src:schemes}.

\section{Communication Pattern}\label{src:apps}
This section describes the various communication patterns addressed by the
different GPU-centric schemes. The different communication
patterns are broadly grouped into three categories: (1) nearest-neighbor
communication, (2) data-driven communication, and (3) persistent collective
communication. This section does not describe the API syntax and semantics
required to implement the GPU-centric communication schemes. Instead, it
provides high-level data movement requirements of some prominent HPC and ML
applications that could adopt GPU-centric communication schemes.

\subsection{Nearest-neighbor Communication}\label{src:nearest-neighbor}
Near-neighbor communication appears in many contemporary HPC simulation
applications. It represents one of the most essential communication patterns,
primarily using two-sided point-to-point communication operations like
\textit{send} and \textit{receive} operations supported by MPI~\cite{specmpi}.
The communication data in this pattern is usually packed at the source process
and sent to the target process, where the received data is unpacked before
consumption. There is usually, at most, one message in the communication pairs
for each phase, and there can be multiple neighbors per process in each phase.

With GPU-attached memory buffers associated with the communication, a GPU
compute kernel performs the packing and unpacking operations, and the
communication operations are performed at GPU kernel boundaries. Some prominent
examples of this communication pattern in HPC simulation frameworks include
adaptive mesh refinement frameworks like BoxLib~\cite{boxlib}, Navier-Stokes
CFD solvers like Nekbone~\cite{Nekbone}, and fusion simulation applications
like Princeton Gyrokinetic Toroidal Code (GTC-P)~\cite{gtc-p}.

\subsection{Data-driven Communication}\label{src:data-driven}
Large-scale data analytics problems like sorting require finding meaningful
patterns in data sets. Effective solutions for addressing such problems
generate fine-grained data movement patterns between unpredictable sets of
processes. These resulting irregular communication patterns are different from
the relatively fixed communication patterns observed in traditional simulation
workloads~\cite{Nekbone,gtc-p,boxlib} (as discussed in
section~\ref{src:nearest-neighbor}) that exploits the structural regularity and
the data locality in the problem set.

While message passing~\cite{specmpi} is established as a \textit{de facto}
programming style for addressing the communication requirements for the HPC
simulation workloads, Partitioned Global Address Space
(PGAS)~\cite{pgasspec} style of programming is considered a highly effective
~\cite{sc14genome,parallelHessianGlobalUpdates,naturalGrainedMultiGridSolver,
matrixSamplingGNN,MetaHipMer,MerBench,asyncTaskCholeskySolver} alternate for addressing the communication
requirements of data-driven workloads. Enabling GPU-centric communication
schemes with PGAS-style of programming allows enabling the data-driven
communication operations to exploit using these advanced communication
operations involving GPU-attached memory buffers.

\subsection{Collective Communication}
While sections~\ref{src:nearest-neighbor} and ~\ref{src:data-driven} described
contrasting communication patterns (regular vs.\ irregular), both represented a
pairwise exchange model where the associated operations are between a pair of
processes. This section describes the communication operations with
GPU-attached memory buffers performed by a group of processes.

While collective communication operations are widely used on HPC and ML
workloads~\cite{gi-fine-grain-overlap-hamodoche}, this section provides examples of their use in distributed ML
training and inference applications. Distributed ML workloads employ different
parallelism~\cite{tal-ben-ddnn} strategies, such as tensor parallelism, data
parallelism, layer parallelism, and hybrid parallelism, that generate data
movement operations involving wide use of collective communication operations.
Examples of such operations include \textit{allreduce},
\textit{reduce-scatter}, \textit{allgather}, and \textit{broadcast}. The type
of collective communication depends on the parallelism strategy. Enabling
GPU-centric communication schemes as collective communication operations can
enable the effective execution of distributed ML workloads.

\subsection{Communication Pattern Comparison}
Section~\ref{src:apps} introduced three application groups representing widely used communication patterns across HPC and ML applications employing GPU-attached memory buffers in the data movement operations. Table~\ref{tab:comm-pattern} provides a rough representation of the various traits of communication patterns discussed in this section. Based on our understanding, the potential GPU-centric schemes associated with each communication pattern are briefly mentioned in the table. However, details on the message payload size and the frequency of the operation are not discussed as these are application-dependent details.

\begin{table*}[]
\centering
\caption{Comparing communication patterns addressed by GPU-centric communication schemes.}
\label{tab:comm-pattern}
\begin{tabular}{|l|l|l|l|}
\hline
                  & \textbf{Nearest-Neighbor} & \textbf{Data-driven} & \textbf{Collective} \\ \hline
Pattern           & Regular                   & Irregular / Random   & Regular             \\ \hline
Execution         & Kernel boundaries         & Persistent Kernels   & Kernel boundaries   \\ \hline
Pairs             & Pairwise                  & Pairwise             & Group-of-processes  \\ \hline
Protocols         & Point-to-point            & RMA                  & Collectives         \\ \hline
Style             & Message Passing           & PGAS                 & NA                  \\ \hline
Examples          & HPC Simulation            & Data Analytics       & Distributed ML      \\ \hline
Potential Schemes & ST                        & KI                   & ST / KI             \\ \hline
\end{tabular}
\end{table*}

Section~\ref{src:apps} introduced three widely used communication patterns in
HPC and ML workloads that can potentially employ GPU-centric communication
schemes. While the communication pattern described in this section represents
some critical use cases, it does not cover all the known use cases. This
section can be considered a high-level representation of the potential use
cases that can eventually be addressed by the GPU-centric communication schemes
exhibited in section~\ref{src:schemes}. Section~\ref{src:challenges} describes
the various challenges in implementing the discussed schemes.

%% file: src/content/challenges.tex
\section{Challenges and Open Questions}\label{src:challenges}

This section outlines the central challenges and open questions in the field
of GPU-centric communication to inspire future research.

\begin{enumerate}
\item \textbf{Efficient compute node architecture}. While some of the features
factored into the compute node design, such as the GPU-to-NIC ratio, GPU-to-CPU
link (PCIe vs.\ proprietary link like NVlink and Infinity fabric), link
connection (CPU vs.\ GPU), number of NICs per node, and number of GPUs per node
are discussed in section~\ref{src:systems}, how to indentify an efficient
compute node architecture for exposing GPU-centric communication? This requires
further analysis of the various compute node components impacting the
performance of GPU-centric communication.

\item \textbf{Communication protocol selection}. While the requirements of the
GPU-aware applications provide details on the preferred communication protocol
(message passing-based synchronous vs.\ one-sided RMA-based asynchronous), it
is not clear on the supportability of the GPU-centric communication schemes in
the existing communication protocols. What is the effective communication
protocol for exposing different discussed GPU-centric communication schemes?

\item \textbf{Standardization.} The adaptability of the GPU-centric
communication schemes depends on the potential to standardize the operations
based on these schemes into specifications such as MPI~\cite{specmpi} or
OpenSHMEM~\cite{specosm}. Vendor-specific libraries exposing these
communication schemes hinder portability and broad adoption. How to
standardize the syntax and semantics of the operations exposing the GPU-centric
communication? Is there specific programming model to target for
standardization?

\item \textbf{NIC feature requirements}. Hardware co-design with respect to the
network interconnect design is critical for enabling a performant implementation
of the data movement operation representing the GPU-centric communication
scheme. How to co-design NIC features with the operations exposing
GPU-centric communication schemes?

\item \textbf{GPU device requirements}. Similar to previous open question, how
to co-design GPU architectures with the data movement operations exposing
GPU-centric communication schemes?

\item \textbf{Software layer support}. With the GPU-centric communication
support relatively recent, communication middlewares (like Libfabric, UCX, and
verbs) required to expose the NIC hardware features are not designed to support
these new schemes. Is it possible to integrate these new schemes into the
existing middleware libraries or do we need separate libraries for these
advanced GPU-aware schemes?

\item \textbf{Application adaptability}. Most applications are designed to be 
GPU-aware. However, adapting to the new GPU-centric communication schemes might 
require extensive code change (like moving from explicit GPU kernels to long-
running persistent GPU kernels). How can applications be created to use the 
communication operations supporting GPU-centric communication schemes?

\end{enumerate}

%% file: src/content/conclusion.tex
\section{Concluding Remarks}\label{src:conclude}

With GPUs identified as a prominent compute node component in modern
heterogeneous supercomputing systems, it is imperative to determine efficient
communication operations associated with GPU-attached memory buffers. The
existing state-of-the-art approach, GPU-aware communication, requires a CPU to
orchestrate the data movement operations associated with the GPU-attached
memory buffers. GPU-centric communication is a relatively new approach that
offloads the control path of the communication operation from the CPU to the
GPU. This work exhibits the various available GPU-centric communication
schemes. We discuss the need for these new communication schemes, factors
impacting the implementation of these schemes, and potential communication
patterns addressed by these new schemes. With multiple standards committees (
like MPI and OpenSHMEM) and user communities interested in introducing the data
movement operations exposing GPU-centric communication schemes, this work
provides a detailed description of these communication schemes. It could
inspire future research to address the known open challenges.